\documentclass[a4paper,12pt]{article}
\pdfoutput=1
\usepackage{graphicx,rotating,hyperref,slashed,amsmath,xcolor,amssymb,amsfonts,expdlist,colortbl,cite,youngtab}
\usepackage{multirow}
\usepackage{booktabs}    
\usepackage{calc}        
\makeatletter
\hypersetup{colorlinks,bookmarksopen,bookmarksnumbered,
linkcolor=blus,pdfstartview=FitH,urlcolor=rossos,citecolor=verde}

\def\lsim{\mathrel{\rlap{\lower3pt\hbox{\hskip0pt$\sim$}}
   \raise1pt\hbox{$<$}}}         
\def\gsim{\mathrel{\rlap{\lower4pt\hbox{\hskip1pt$\sim$}}
   \raise1pt\hbox{$>$}}}         

 \newcommand{\sfootnote}[1]{}
\definecolor{bluc}{cmyk}{1,1,0,0.1}
\definecolor{rossoCP3}{cmyk}{0,.88,.77,.40}
\definecolor{rosso}{cmyk}{0,1,1,0.4}
\definecolor{rossos}{cmyk}{0,1,1,0.55}
\definecolor{rossoc}{cmyk}{0,1,1,0.2}
\definecolor{verdes}{cmyk}{0.92,0,0.59,0.4}

\hypersetup{colorlinks, bookmarksopen, bookmarksnumbered,
citecolor=verdes, linkcolor=bluc, pdfstartview=FitH, urlcolor=rossos}

\newcommand{\mio}[1]{}

\definecolor{Gray}{gray}{0.95}

\newcommand{\sfrac}[2]{#1/#2}

\usepackage{multicol}
\usepackage{color}
\definecolor{rosso}{cmyk}{0,1,1,0.4}
\definecolor{rossos}{cmyk}{0,1,1,0.55}
\definecolor{rossoc}{cmyk}{0,1,1,0.2}
\definecolor{blu}{cmyk}{1,1,0,0.3}
\definecolor{blus}{cmyk}{1,1,0,0.6}
\definecolor{bluc}{cmyk}{1,1,0,0.1}
\definecolor{verde}{cmyk}{0.92,0,0.59,0.25}
\definecolor{verdec}{cmyk}{0.92,0,0.59,0.15}
\definecolor{verdes}{cmyk}{0.92,0,0.59,0.4}

\renewcommand\_{_}

\newcommand{\eq}[1]{~{\rm (\ref{eq:#1})}}

\newcommand{\GeV}{\,{\rm GeV}}
\newcommand{\TeV}{\,{\rm TeV}}

\newcommand{\Tr}{\,{\rm Tr}}

\def\circa#1{\,\raise.3ex\hbox{$#1$\kern-.75em\lower1ex\hbox{$\sim$}}\,}

\newcommand{\beq}{\begin{equation}}
\newcommand{\eeq}{\end{equation}}

\newcommand{\bea}{\begin{eqnarray}}
\newcommand{\eea}{\end{eqnarray}}
\newcommand{\be}{\begin{equation}}
\newcommand{\ee}{\end{equation}}
\font\tenrsfs=rsfs10 at 12pt
\font\sevenrsfs=rsfs7 at 10 pt
\font\fiversfs=rsfs5
\newfam\rsfsfam
\textfont\rsfsfam=\tenrsfs
\scriptfont\rsfsfam=\sevenrsfs
\scriptscriptfont\rsfsfam=\fiversfs
\def\mathscr#1{{\fam\rsfsfam\relax#1}}

\def\Lag{\mathscr{L}}

\def\circa#1{\,\raise.3ex\hbox{$#1$\kern-.75em\lower1ex\hbox{$\sim$}}\,}
\makeatletter

\def\hhref#1{\href{http://arxiv.org/abs/#1}{arXiv:#1}} 

\newcommand{\doi}[1]{\href{http://dx.doi.org/#1}{[doi]}}

\def\hhref#1{\href{http://arxiv.org/abs/#1}{arXiv:#1}} 
 
\def\art{\@ifnextchar[{\eart}{\oart}}
\def\eart[#1]#2#3#4#5#6{{\rm #2}, {\em #3 \bf #4} {\rm (#6) #5} ({\em #1})}

\def\article{\@ifnextchar[{\earticle}{\oarticle}}
\def\oarticle#1#2#3#4#5#6{{\rm #1}, {\em ``#6''}, {\rm #2 #3 (#5) #4}}
\def\earticle[#1]#2#3#4#5#6#7{{\rm #2}, {\em ``#7''}, {\rm #3 #4 (#6) #5}  [\hhref{#1}]}
\def\hepart[#1]#2{{\rm #2, \em#1}}
\def\heparticle[#1]#2#3{#2, {\em ``#3''} [\hhref{#1}]}

\newcounter{alphaequation}[equation]
\def\thealphaequation{\theequation\hbox to
0.6em{\hfil\alph{alphaequation}\hfil}}
\def\eqnsystem#1{
\def\@eqnnum{{\rm (\thealphaequation)}}
\def\@@eqncr{\let\@tempa\relax \ifcase\@eqcnt \def\@tempa{& & &} \or
  \def\@tempa{& &}\or \def\@tempa{&}\fi\@tempa
  \if@eqnsw\@eqnnum\refstepcounter{alphaequation}\fi
\global\@eqnswtrue\global\@eqcnt=0\cr}
\refstepcounter{equation} \let\@currentlabel\theequation \def\@tempb{#1}
\ifx\@tempb\empty\else\label{#1}\fi
\refstepcounter{alphaequation}
\let\@currentlabel\thealphaequation
\global\@eqnswtrue\global\@eqcnt=0 \tabskip\@centering\let\\=\@eqncr
$$\halign to \displaywidth\bgroup \@eqnsel\hskip\@centering
$\displaystyle\tabskip\z@{##}$&\global\@eqcnt\@ne
\hskip2\arraycolsep\hfil${##}$\hfil& \global\@eqcnt\tw@\hskip2\arraycolsep
$\displaystyle\tabskip\z@{##}$\hfil
\tabskip\@centering&\llap{##}\tabskip\z@\cr}
\def\endeqnsystem{\@@eqncr\egroup$$\global\@ignoretrue} \makeatother

\newcommand{\SU}{\,{\rm SU}}
\newcommand{\SO}{\,{\rm SO}}

\newcommand{\U}{\,{\rm U}}
\newcommand{\N}{N_c}

\definecolor{fiorentina}{rgb}{.5,0,.5}

\allowdisplaybreaks

\begin{document}
\centerline{CERN-{TH}-2017-001
\hfill  CP3-Origins-2017-001 \hfill IFUP-TH/2017}

\vspace{1truecm}

\begin{center}
\boldmath

{\textbf{\LARGE\color{rossoCP3} Naturalness of asymptotically safe Higgs}}
\unboldmath

\bigskip\bigskip

\large

{\bf Giulio Maria Pelaggi}$^a$,
{\bf Francesco Sannino$^b$, \\
Alessandro Strumia$^{a,c}$, Elena Vigiani$^a$}
 \\[8mm]
{\it $^a$ Dipartimento di Fisica dell'Universit{\`a} di Pisa and INFN, Italy}\\[1mm]
{\it $^b$ CP$^3$-Origins and Danish IAS, University of Southern Denmark, 
Denmark}\\[1mm]
{\it $^c$ CERN, Theory Division, Geneva, Switzerland}\\[1mm]

\bigskip\bigskip

\thispagestyle{empty}\large
{\bf\color{blus} Abstract}
\begin{quote}
We extend the list of theories featuring a rigorous interacting ultraviolet fixed point by constructing the first theory featuring a Higgs-like scalar
with gauge, Yukawa and quartic interactions. We show that the theory enters a perturbative asymptotically safe regime  at energies above a  physical scale $\Lambda$.  We determine the salient properties of the theory and use it as a concrete example to test whether scalars masses unavoidably receive
quantum correction of order $\Lambda$.  Having at our dispose a calculable model allowing us to precisely relate the IR and UV of the theory we demonstrate that the scalars can be lighter than $\Lambda$.  
Although we do not have an answer to whether the Standard Model hypercharge coupling growth towards a Landau pole at around $\Lambda \sim 10^{40}\GeV$  can be tamed by non-perturbative asymptotic safety, our results indicate that such a possibility is worth exploring. In fact, if successful, it might also offer an explanation for the unbearable lightness of the Higgs.
\end{quote}
\thispagestyle{empty}
\end{center}

\setcounter{page}{1}
\setcounter{footnote}{0}



\newpage

\section{Introduction}

The Large Hadron Collider (LHC) data at $\sqrt{s}=13\TeV$ confirm the Standard Model (SM) and give strong bounds on supersymmetry, on composite Higgs and on other SM extensions that were put forward to tame the 
 quadratically divergent corrections to the Higgs mass in a natural way. The existence of natural solutions apparently  ignored by nature challenges even anthropic approaches.  This unsettling situation calls for reconsidering the issue of naturalness.

\smallskip

The bulk of the physical corrections to the SM observables are only logarithmically sensitive to a potential UV physical scale because they stem from marginal operators. Physical corrections to the Higgs mass are  small in the SM,
and can remain small once it is extended to account for
dark matter, neutrino masses~\cite{1303.7244}, gravity, and inflation~\cite{agravity}. This is true up to possible power-divergent corrections  that may  offset the lightness of the Higgs. 
As well known, the Higgs propagator $\Pi(q^2)$
at zero momentum $q=0$ receives a 
quadratically divergent correction, which is often interpreted as a large correction to the Higgs mass.
Writing only the top Yukawa one-loop contribution, one has
\beq \label{eq:0}
 \Pi(0)
 =-12 y_t^2  \frac{1}{i}\int \frac{d^4k}{(2\pi)^4} \frac{ k^2 + m_t^2}{(k^2-m_t^2)^2}+\cdots
\eeq
The photon too receives at zero momentum a quadratically divergent correction.  In QED one has
\beq\label{eq:1}
\Pi_{\mu\nu}(0) = -4 e^2 \frac{1}{i} \int \frac{d^4k}{(2\pi)^4} 
\bigg[ \frac{2k_\mu k_\nu}{(k^2-m_e^2)^2} - 
\frac{\eta_{\mu\nu}}{k^2-m_e^2}\bigg].\eeq
This is not interpreted
as a large photon mass because it is presumed that 
some unknown physical cut-off regulates divergences while
respecting gauge invariance, that forces the photon to be massless.
Similarly, the graviton propagator receives a quadratically divergent correction $\Pi_{\mu\nu,\rho\sigma}(0)$:
in part it can be interpreted as a cosmological constant, in part it breaks reparametrization invariance. 

\medskip

The fate of the Higgs mass is not clear.
Some regulators (such as dimensional regularization) respect all these symmetries and
get rid of all power divergences, including the one that affects the Higgs mass. 
Other regulators (such as Pauli-Villars and presumably string theory) 
do not generate a photon mass nor a graviton mass and
generate a large Higgs mass, 
given that it is only protected by scale invariance,
which is not a symmetry of the full theory.


\begin{figure}[t]
$$\includegraphics[width=\textwidth]{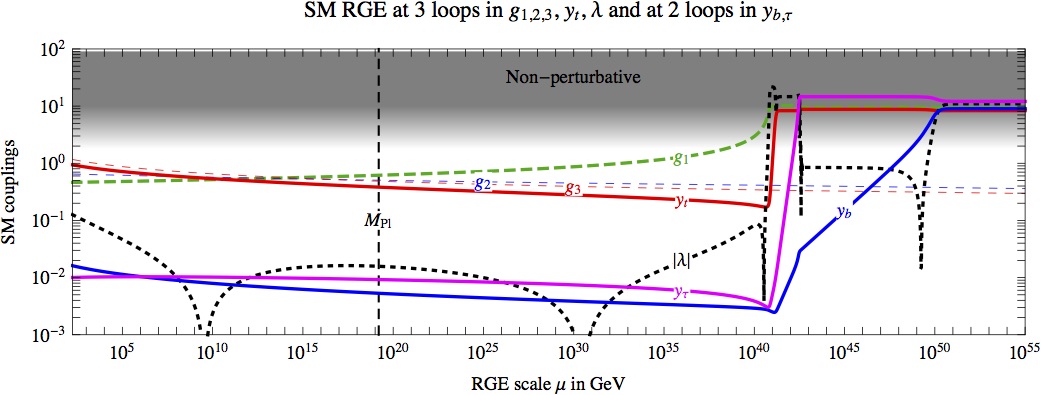}$$
\caption{\em Illustration of a possible RGE running in the SM.
We assumed central values for all parameters, and solved the 3 loop RGE equations.
In order to obtain an asymptotically safe behaviour 
we artificially removed the bottom and tau Yukawa contributions to the 3-loop term in the RGE.
This only affects the running in the non-perturbative region above $10^{40}\GeV$, 
where the result cannot of course be trusted.
Furthermore we ignored the Yukawa couplings of the lighter generations, and gravity.
\label{fig:SMRGE}
}
\end{figure}

\bigskip

One possibility is that the SM is (part of) a theory valid up to infinite energy,  such that no 
physical cut-off exists.
Then, once that eq.\eq{1} is interpreted to mean zero, the same divergence in eq.\eq{0} must be interpreted in the same way.
Furthermore, in a theory with dimension-less parameters only, one can argue that $\int d^4 k/k^2 = 0$ by dimensional analysis.
Gravity itself could be described by small dimension-less parameters~\cite{agravity,TAF},
such that it makes sense to extrapolate the SM RGE above the Planck scale.

In this context, one possibility is devising
realistic weak-scale extensions of the SM such that  all gauge, Yukawa, and quartic
couplings  flow to zero at infinite energy~\cite{TAF}.
However
hypercharge must be embedded into a large non-abelian group, in order to be asymptotically free:
naturalness then demands new vectors at the weak scale, which have not been observed so far.

The other possibility is that the SM itself might be asymptotically safe.
The hypercharge gauge coupling $g_Y$ becomes non-perturbative at $\Lambda \sim 10^{40}\GeV$, hitting a `Landau pole'. 
It is not known what it means. 
It might mean that the SM is not a complete theory and new physics is needed at lower energy.
Otherwise $g_Y$ and other couplings might run up to constant non-perturbative values
as illustrated in fig.~\ref{fig:SMRGE},
such that the SM enters into an asymptotically safe phase. In fact, this possibility was envisioned very early on in the literature \cite{GellMann:1954fq,Weinberg} triggering lattice studies  \cite{Gockeler:1997dn,Kim:2000rr,Kim:2001am} as well as  non-perturbative analytic studies such as the one of \cite{Gies:2004hy}.  It is fair to say, however, that the fate of the SM depends on non-perturbative effects which are presently unknown;
see~\cite{Bogoliubov,redmond,hep-ph/9704333,0804.2650,1010.4081,1011.0061}
for attempts to compute the non-perturbative region and for related ideas. 
 
\medskip

Tavares, Schmaltz and Skiba~\cite{Skiba} proposed an alleged no-go argument,
according to which {\it Landau makes Higgs obese}: i.e. scalars generally receive a mass correction  of the order of the would-be-Landau pole scale $\Lambda$.
In the SM case, this would mean that, whatever happens at $10^{40}\GeV$,
the Higgs mass receives a contribution of order $10^{40}\GeV$,
so that an asymptotically safe Higgs (where asymptotic safety kicks above $\Lambda$) cannot be natural.

Later, Litim and Sannino (LS)~\cite{SL} presented the first four-dimensional example of a perturbative quantum field theory
where {\it all} couplings that are small at low energy flow to a constant value at higher energy persisting up to infinite energy. This model involves a gauge group $\SU(\N)$ with large $\N$,
a neutral scalar $S$ and vector-like charged fermions, with asymptotically safe
Yukawa couplings and scalar quartics. The model realises Total Asymptotic Safety (TAS). Another equally relevant property of the model is that without the scalar it cannot be perturbatively safe \cite{Caswell:1974gg,SL}. Scalars are required to dynamically render the theory fundamental at all scales without invoking supersymmetry,
which would keep scalars massless independently of their dynamics.\footnote{It is important to note that scalars without the simultaneous presence of gauge and fermion interactions, are not ultraviolet safe as a large body of analytic and first principle lattice results has demonstrated \cite{Aizenman:1981du,Frohlich:1982tw,Luscher:1987ay,Luscher:1987ek,Heller:1992pj,Wolff:2009ke,Korzec:2015pma}.} In fact supersymmetry makes it harder to realise an asymptotically safe scenario \cite{Intriligator:2015xxa,Bajc:2016efj} both perturbatively and non-perturbatively.
Furthermore the LS model, on the line of physics, connects two fixed points, a non-interacting infrared free one (the theory at low energy is non-abelian QED-like) to an interacting ultraviolet fixed point. Remarkably the model shares the SM backbone since it features gauge, fermion and needed scalar degrees of freedom, albeit it still misses a gauged Higgs-like state.  
{ We therefore extend the LS model in section~\ref{LS+} to further feature a Higgs-like charged scalar $H$.  We  rigorously demonstrate  that the theory enters a perturbative asymptotically safe regime  at energies above a  physical scale $\Lambda$. We also show that we can determine the RGE flow linking ultraviolet and infrared physics precisely. 
}

{ In the appendix we explore theories featuring chiral fermions and show that it is possible to achieve asymptotic safety for the gauge and Yukawa couplings while safety for scalar couplings is challenging.} 

Having at our disposal a calculable model similar to the SM, we carefully re-consider the naturalness issue in this class of theories in order to offer an answer to the question: Does the Higgs-like scalar $H$ acquire a mass of the order of the scale $\Lambda$?
In section~\ref{H} we do not find any such contribution, de facto, re-opening the issue. We discuss possible caveats and offer our conclusions in section~\ref{concl}.

\section{Asymptotically safe models with an Higgs-like  scalar}\label{LS+}
Litim and Sannino considered a model with gauge group $\SU(N_c)$ and gauge coupling $g$;
$N_F$ vector-like fermions $\psi_i\oplus\bar\psi_i$ in the fundamental plus anti-fundamental,
and $N_F^2$ neutral scalars $S_{ij}$ with Yukawa couplings
$S_{ij} \psi_i \bar\psi_{j}$.
The number of flavours $N_F$ can be fixed to make the one-loop gauge beta function $\beta_g^{(1)}$ small.
Large $\N$, $N_F$ allows to make $\beta_g^{(1)}$ arbitrarily small, guaranteeing perturbative control.
The new key feature with respect to the analogous construction by Banks and Zaks~\cite{BZ}
is that the Yukawa couplings can (non trivially) make the two-loop gauge beta function negative \footnote{ It was indeed shown for the first time in \cite{Antipin:2013pya,SL,Litim:2015iea} that Yukawa interactions are instrumental, in perturbation theory, to tame the UV behaviour of non asymptotically free gauge theories. In fact without scalars gauge-fermion theories, 
in perturbation theory, are doomed to remain at best effective field theories \cite{Caswell:1974gg,SL}. These conditions were further elaborated in \cite{Bond:2016dvk} and in \cite{Molgaard:2016bqf} for chiral matter.  Of course having a fixed point in the Yukawa and gauge coupling is not enough for the theory to be safe, and much more work is required to show that it is safe in all couplings. Beyond perturbation theory one can argue that at large number of matter flavours and finite number of colours one can achieve asymptotic safety as recently summarised and further elucidated in \cite{Antipin:2017ebo}. Exact non perturbative results have been established for supersymmetric field theories \cite{Intriligator:2015xxa}.  },
such that, together with $\beta_g^{(1)}>0$,  $g$ enters into a perturbative fixed point at large energy.
Finally, one verifies that Yukawa couplings and scalar quartics too have a perturbative fixed point.
The model satisfies Total Asymptotic Safety (TAS).

In general the equations $\beta_g=\beta_y = \beta_\lambda=0$  have multiple solutions, 
that correspond to different global symmetries of the theory.
The analysis can be simplified focusing on the maximal global symmetry, $\U(N_F)_L\otimes\U(N_F)_R$,
which can be realized with complex scalars $S$. 
The field content is then summarized by the upper box of table~\ref{tab:SL-chargedscalar}.

\begin{table}\small
$$\begin{array}{|c|cc|cc|}\hline
\hbox{Fields}& \multicolumn{2}{c|}{\hbox{Gauge symmetries}} & \multicolumn{2}{c|}{\hbox{Global symmetries}} \\
\hline  
    & \hbox{Spin}  & \SU(\N)  & \U(N_F)_L & \U(N_F)_R   \\ 
\hline  \hline
\rowcolor[cmyk]{0.1,0,0,0}  \psi & 1/2 & {\tiny  \yng(1)} & \overline{\tiny  \yng(1)} & 1   \\
\rowcolor[cmyk]{0.1,0,0,0}  \bar\psi & 1/2&\overline{\tiny  \yng(1)} & 1 & {\tiny  \yng(1)}   \\
\rowcolor[cmyk]{0.1,0,0,0}  S &0& 1 & {\tiny  \yng(1)} & \overline{\tiny  \yng(1)}  \\ 
\hline \hline 
\rowcolor[cmyk]{0,0.1,0,0}  H &0& {\tiny  \yng(1)} & 1& 1  \\ 
\rowcolor[cmyk]{0,0.1,0,0}  N & 1/2 & 1 & 1 & \overline{\tiny  \yng(1)}   \\
\rowcolor[cmyk]{0,0.1,0,0}  N' & 1/2 &  1 & {\tiny  \yng(1)} & 1  \\
\hline
\end{array}$$
\caption{\em\label{tab:SL-chargedscalar}  Field content of the model.
The upper box is the original Litim-Sannino model. 
The lower box are the extra fields that we add in order to get a Higgs-like scalar $H$.}
\end{table}

The lower box of table~\ref{tab:SL-chargedscalar} shows the fields that we add: one Higgs-like scalar charged
under the gauge group.
Its introduction does not 
affect, in the limit of large $N_c$, $N_F$, the fixed point for $y$ and $g$ found in \cite{SL}. 
We also add singlet fermions $N_i$, $N'_i$ (see table~\ref{tab:SL-chargedscalar} for the details) 
in order to allow $H$ to have Yukawa couplings, like the SM Higgs.
The allowed Yukawa couplings then are
\beq\label{eq:LagY}
\Lag_Y = y \, S_{ij } \psi_i \bar{\psi}_j +y'\, S_{ij}^* N_i N'_j+
 \tilde{y} \,H \bar\psi_i N_i + \tilde{y}' \,H^* \psi_i N'_i + {\rm h.c.}  
\eeq
The scalar potential is
\beq
V = 
 \lambda_{S1} (\!\Tr S^\dag S)^2 + \lambda_{S2} \Tr(S^\dag S S^\dag S) +  \lambda_{H} (H^\dag H)^2  
+ \lambda_{HS}(H^\dag H) \Tr(S S^\dag)    \,,
\eeq
and it is positive if
\bea \label{eq:Vpos}
\lambda_{S1} + \eta \lambda_{S2}  \geq 0 \,, \quad \lambda_H > 0 \,, \quad \lambda_{HS} + 2 \sqrt{\lambda_H (\lambda_{S1} + \eta \lambda_{S2})} \geq 0 \,,
\eea
where $\eta = \sfrac{\Tr( S^\dag S S^\dag S )}{\Tr^2(S^\dag S)}$ ranges between $\eta = 1$ and $\eta = 1/N_F$.
The bounds in eq.\eq{Vpos} need only to be imposed at the extremal values.

\subsection{RGE and their fixed points}
Defining the $\beta$-functions coefficients as 
\beq
\frac{d X}{d \ln \mu} = \frac{\beta_X^{(1)}}{(4\pi)^2} +  \frac{\beta_X^{(2)}}{(4\pi)^4} + \dots \,, 
\eeq
the relevant RGE are  
\begin{eqnsystem}{sys:RGE+}
\beta_{g}^{(1)}  &=&  g^3 \left(-\frac{11 }{3}N_c +\frac{2 N_F}{3}+\frac{1}{6}\right) \\
\beta_{g}^{(2)}  &=&  g^5 \left(\frac{13 N_c N_F}{3}-\frac{N_F}{N_c}-\frac{34 N_c^2}{3}+\frac{4 N_c}{3}-\frac{1}{N_c}\right) - g^3 \left(N_F^2 y^2 +  N_F \frac{\tilde{y}^2+ \tilde{y}'^2 }{2}\right)  \\
\beta_{y}^{(1)}  &=&  y^3 \left(N_c+N_F\right)+g^2 y \left(\frac{3}{N_c}-3 N_c\right)+y  \frac{\tilde{y}^2+ \tilde{y}'^2 }{2}  + y y'^2 + 2 y' \tilde{y} \tilde{y}' \\
\beta_{y'}^{(1)}  &=&  N_c y^2 y' + y'^3(N_F+1)  +   y' N_c \frac{\tilde{y}^2 + \tilde{y}'^2}{2} + 2 N_c y \tilde{y} \tilde{y}'  \\
\beta_{\tilde{y}}^{(1)}  &=&  N_F \tilde{y}  \frac{y^2+y'^2+2 \tilde{y}'^2}{2}  +\tilde{y}^3 \left(\frac{N_c}{2}+N_F+\frac{1}{2}\right)+g^2 \tilde{y} \frac32 \frac{1-\N^2}{\N}  + 2 N_F y y' \tilde{y}'  \\
\beta_{\tilde{y}'}^{(1)}  &=&  N_F \tilde{y}' \frac{ y^2+ y'^2+2\tilde{y}^2}{2} +\tilde{y}'^3 \left(\frac{N_c}{2}+N_F+\frac{1}{2}\right)+g^2 \tilde{y}'
\frac32 \frac{1-\N^2}{\N}  + 2 N_F y y' \tilde{y} \\
\beta_{\lambda_{S1}}^{(1)}  &=&  4 y^2 N_c \lambda _{{S1}} + 4 y'^2 \lambda _{{S1}} +N_c \lambda _{{HS}}^2+\left(4 N_F^2+16\right) \lambda _{{S1}}^2+16 N_F \lambda _{{S1}} \lambda _{{S2}}+12 \lambda _{{S2}}^2  \\
\beta_{\lambda_{S2}}^{(1)}  &=&  4 y^2 N_c \lambda _{{S2}}-2 y^4 N_c+8 N_F \lambda _{{S2}}^2+24 \lambda _{{S1}} \lambda _{{S2}} - 2 y'^4 + 4 y'^2 \lambda_{S2}  \\
\beta_{\lambda_{H}}^{(1)}  &=&  g^4 \left(\frac{3 N_c}{4}-\frac{3}{N_c}+\frac{3}{2 N_c^2}+\frac{3}{4}\right)+g^2 \left(\frac{6}{N_c}-6 N_c\right) \lambda _{{H}}+\left( 4  N_c+16\right) \lambda _{{H}}^2+ \notag \\
&&+4 N_F \lambda _{{H}} (\tilde{y}^2+ \tilde{y}'^2)+N_F^2 \lambda_{{HS}}^2-2 N_F( \tilde{y}^4+ \tilde{y}'^4 )\\
\beta_{\lambda_{HS}}^{(1)}  &=&  2 (y^2 N_c+ y'^2 ) \lambda _{{HS}}+(\tilde{y}^2+\tilde{y}'^2) \left(2 N_F \lambda _{{HS}} -4 y^2 -4 y'^2\right)+g^2  \lambda _{{HS}}\left(\frac{3}{N_c}-3 N_c\right)+ \notag \\
&&+\left(4 N_c+4\right) \lambda _{{H}} \lambda _{{HS}}+\lambda _{{HS}} \left(\left(4 N_F^2+4\right) \lambda _{{S1}}+8 N_F \lambda _{{S2}}\right)+4 \lambda _{{HS}}^2 - 8 y y' \tilde{y} \tilde{y}' 
\end{eqnsystem}
Notice that $y y' \tilde{y}^* \tilde{y}^{\prime *}$  is left invariant by redefinitions of the phases of all fields, so the model admits one CP-violating phase.
Nevertheless CP is conserved at all fixed points, so that the RGE can be written in terms of real couplings.
 For simplicity we therefore assume all couplings to be real.

The one-loop gauge beta function can be rewritten as
\beq \beta_{g}^{(1)} =  g^3 \frac{2\N}{3}\epsilon,\qquad\hbox{where}\qquad
\epsilon \equiv   \frac{N_F}{\N}-\frac{11}{2} +\frac{1}{4N_c}\eeq
can be made arbitrarily small in the limit of large $N_c,N_F$.
In this limit $\beta_{y}^{(1)} $ reduces to
\beq 
\beta_{y}^{(1)} \stackrel{\N\gg 1}{\simeq}  \N y \bigg(-3 g^2 + \frac{13}{2} y^2\bigg)\eeq
and it vanishes for $y^2/g^2 \simeq 6/13$, which corresponds to a negative
\beq \label{eq:betag2}
\beta_g^{(2)} {\simeq} \frac{25}{2}  g^5 \N^2 \bigg( 1 - \frac{363}{325}\bigg).\eeq
Thereby the gauge coupling has an IR-attractive fixed point $g=0$ and a non-trivial UV-attractive
fixed point at
\beq
\label{gFP-SL}
g^2  = g_*^2 \simeq \frac{26(4\pi)^2}{57 \N} \epsilon \,.
\eeq
The scalar quartics $\lambda_{S1},\lambda_{S2}$ admit
two fixed points. At leading order in $\epsilon$:
\begin{eqnsystem}{sys:FPlamb}
\frac{\lambda_{S1}}{g^2} &\simeq&  
\frac{3}{143 N_F} \left(- 2 \sqrt{23} \pm \sqrt{20 + 6 \sqrt{23}} \right)
\approx
-\frac{1}{N_F}\left\{\begin{array}{ll} 0.348 
 & - \\ 0.055 
  & +\end{array}\right.\\
 \frac{\lambda_{S2}}{g^2} &\simeq & 
\frac{3}{143} \left(\sqrt{23}-1 \right) \approx 0.080 . 
\end{eqnsystem}
The solution with the $+$ ($-$) sign corresponds to a stable (unstable) potential $V(S)$ as determined in \cite{Litim:2015iea}.
At the stable solution, the fixed point for both quartics, as well as the fixed point for $y$,
are IR-attractive: this means that their low-energy values are univocally fixed, with respect to $g$, 
along the RGE trajectory that reaches
infinite energy.

\newcommand{\UV}{{\color{blue}_{\rm UV}}}
\newcommand{\IR}{{\color{red}_{\rm IR}}}

\begin{table}
\begin{center}
\hspace{-30pt}
\begin{tabular}{|c|c|c||c|c|c|c|c|} 

\hline
 $y/g$   & $N_F \lambda_{S1}/g^2$ & $\lambda_{S2}/g^2$ & $\tilde y/g$ & $y',\tilde y'/g$ &
$\lambda_{H}/g^2$ & $N_F \lambda_{HS}/g^2$  & $V$ \\
\hline
\hline 
\multirow{8}{*}{ $\sqrt{\frac{6}{13}}\IR$} &
\multirow{4}{*}{ $ -0.348\UV$} &
\multirow{4}{*}{ $ 0.080\IR$} & 
\multirow{2}{*}{ $ 0\UV $} &
\multirow{2}{*}{ $ 0\UV $} &
$0.138\UV$ & $0\UV$ & unstable\\ \cline{6-8}

&&&&& $1.362\IR$ & $0\UV$ & unstable 
\\  \cline{4-8}

&&&  \multirow{2}{*}{$1/\sqrt{26}\IR$}& \multirow{2}{*}{$0\UV$}  & $0.163\UV$ & $-0.076\UV$ & unstable
\\  \cline{6-8}

&&&&& $1.125\IR$ & $-0.301\UV$ & unstable 


\\ 
 \cline{2-8}

&\multirow{4}{*}{ $ -0.055\IR$} &
\multirow{4}{*}{ $ 0.080\IR$} & 
\multirow{2}{*}{ $ 0\UV $} &
\multirow{2}{*}{ $ 0\UV $} &
$0.138\UV$ & $0\IR$ & $\ge0$
\\  \cline{6-8}

&&&&& $1.362\IR$  & $0\IR$ & $\ge0$ 
\\ \cline{4-8}

&&&  \multirow{2}{*}{$1/\sqrt{26}\IR$}& \multirow{2}{*}{$0\UV$}  & $0.163\UV$ & $0.301\IR$  & $\ge0$
\\ \cline{6-8}

&&&&& $1.125\IR$ & $0.076\IR$ & $\ge0$ 



\\ \cline{1-8}
\end{tabular}
\caption{\em Fixed points at leading order in $\epsilon$.  
The left panel of the table refers to the Litim-Sannino model;
the right panel to the extra couplings.
All fixed points have $g={g_*}\UV$, and the extra
trivial fixed point with $g=0\IR$ is ignored.
The pedix 
$\UV$ denotes an UV-attractive fixed point;
while $\IR$ denotes an IR-attractive fixed point, where the low-energy value of the coupling is fixed.  The equivalent solutions with $\tilde{y}$, $\tilde{y}'$ exchanged are not shown.
\label{tab:FP}}
\end{center}
\end{table}

So far the new fields that we added just acted as spectators.
We must check that they have their own fixed points.
By studying the full equations we find that
the extra Yukawa couplings $y',\tilde y$ and $\tilde y'$ have 3  inequivalent fixed points.
The fixed points with $y'=\tilde y'=0$ lead to fixed points for the quartics, as listed in table~\ref{tab:FP}.
The full potential $V(S,H)$ is stable when $V(S)$ is stable.

\begin{figure}[t]
$$\includegraphics[width=0.7\textwidth]{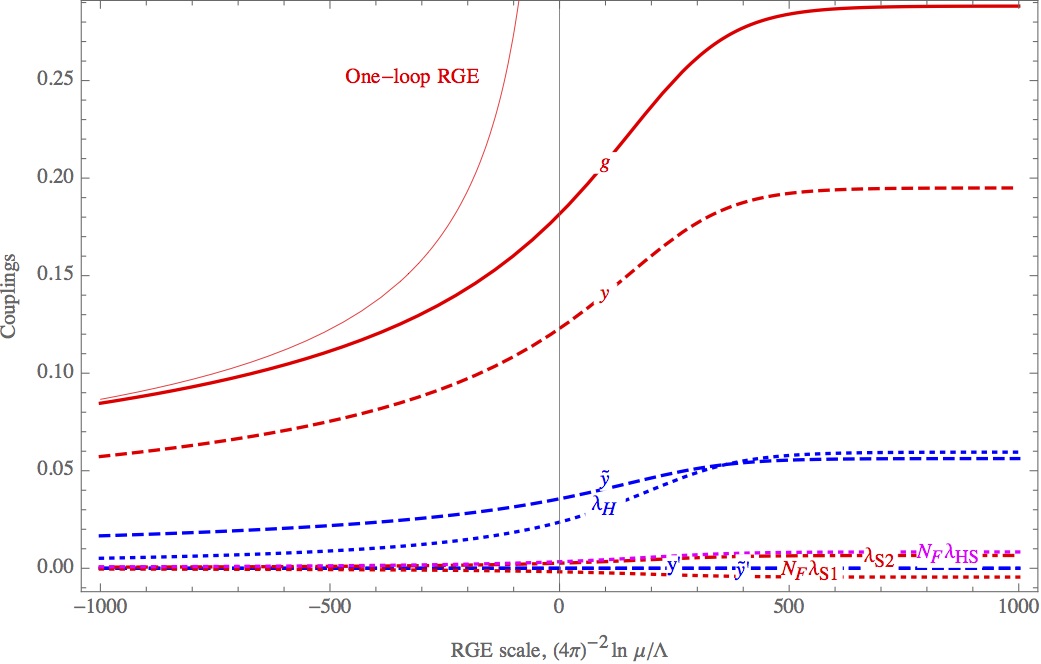}$$
\caption{\em Illustration of a possible RGE running with $\N=10$, $\epsilon=0.01$. \label{fig:modelRGE}
}
\end{figure}

All these couplings are perturbative for $\epsilon \ll 1$, in the sense that higher order corrections are suppressed
by powers of $\epsilon$.
An explicit solution to the RGE equations 
is obtained by assuming that all ratios $y/g$, $\lambda/g^2$ run
remaining constant up to corrections of relative order $\epsilon$.
Then one obtains an  RGE equation for $g$
\beq  \frac{dg}{d\ln\mu}  \stackrel{\epsilon\to 0}{\simeq} \frac{b_1 g^3}{(4\pi)^2}  - \frac{b_2 g^5}{(4\pi)^4} ,\qquad
b_1 =  \frac{2N_c}{3}\epsilon,\qquad b_2 =\frac{19 N_c^2}{13}.\eeq
Its solution is
\beq \ln\frac{\mu}{\Lambda} = -\frac{(4\pi)^2}{2b_1} \bigg[\frac{1}{g^2}+\frac{1}{g_*^2}\ln\bigg((4\pi)^2 b_1 \left(\frac{1}{g^2}-\frac{1}{g_*^2}\right)\bigg)\bigg],
\qquad g^2_* =(4\pi)^2 \frac{b_1}{b_2}.\eeq
which can be used to define in an RGE-invariant way the transmutation scale $\Lambda$
in terms of $\mu$ and of $g(\mu)$.
In the limit where the second two-loop term is neglected, $\Lambda$ becomes the Landau pole scale of one-loop RGE.
Imposing the boundary condition $g(\mu_0)=g_0$ the solution becomes~\cite{SL}
\beq \label{eq:gW}
g^2(\mu)  = \frac{g^2_*}{1+W[({\mu_0}/{\mu})^{2b_1^2/b_2} (g_*^2/g_0^2-1)e^{g_*^2/g_0^2-1} ]}\eeq
where $W(z)$ is the Lambert function defined by $z=We^W$.
The fixed point  $g=g_*$ is UV-attractive: this means that $g$ can become smaller at low energy.
Fig.~\ref{fig:modelRGE} illustrates a typical RGE running.

%
%

\section{On the lightness of safe scalars}\label{H}
We now investigate whether scalars 
can be lighter than the characteristic energy scale $\Lambda$ where the RG flow displays a cross-over from the Gaussian IR scaling to the UV interacting scaling. This scale is dynamical in nature and arises via dimensional transmutation. Above this scale the theory remains finite at arbitrary short distances avoiding the Landau pole. The precise definition and way to determine this scale is  presented in~\cite{Litim:2015iea}.  
 
\medskip
 
The authors of~\cite{Skiba} argued that  scalars acquire masses of order $\Lambda$
by elevating a one-loop computation of the top corrections for the SM Higgs to an operatorial one in an alleged theory featuring an asymptotically safe behaviour. Their rough analysis used: dimensional analysis, modelling the underlying behaviour of the couplings and ad-hoc subtractions to render the result finite. 
Henceforth according to their result the asymptotically safe Higgs scenario would remain unnatural. 
  
 Differently from~\cite{Skiba} we have the precisely calculable model of section~\ref{LS+}, containing a scalar 
$H$ analogous to the Higgs doublet in the SM,
with gauge, Yukawa and quartic interactions, that run  into a perturbative  ultraviolet fixed point
as illustrated in fig.~\ref{fig:modelRGE}. Furthermore the theory connects to a Gaussian IR fixed point. 
We set all masses to zero, making the classical theory scale invariant, and we determine whether quantum corrections make scalars massive. We perform our computations in such a way that both IR and UV quantum conformality are preserved.

\subsection{Perturbative effects}
We start by considering the quantum corrections affecting scalar masses that stem from the standard perturbative approach of summing Feynman diagrams. We therefore fix the renormalisation scale at an arbitrary value $\mu_0$, and compute the scalar two-point function $\Pi(0)$.  At any loop order, each diagram is proportional to powers of dimension-less coupling constants renormalized at $\mu_0$ times a loop integral which is scale-invariant and quadratically divergent
at large loop momenta. To ensure short and large distance quantum conformality these divergences are set to zero. 
As well known, scale-invariant loop integrals vanish automatically in dimensional regularisation.

\subsection{Resumming large logarithms, $\Lambda$ dependence and meaning}\label{RGE}
One might be worried that a mere perturbative analysis is insufficient to settle the issue. 
So, we now comment on potentially different non-perturbative corrections. 

A relevant class of dominant non-perturbative corrections are those where
couplings get enhanced by large logarithms.
At one loop one encounters corrections of relative order
$C\ell$ where
$C =N_c  g^2 /(4\pi)^2$ and $\ell =\ln (E/\mu_0)$.
The correction $C\ell$ becomes of order one at energy $E$ much different from $\mu_0$.
At two loops one encounters corrections of order $C^2\ell^2$ and $C^2 \ell$.

As well known, all corrections of order $(C\ell)^n$ can be resummed by solving the one-loop RGE equations;
all corrections of order $C^n \ell^{n-1}$ are resummed by solving the two-loop RGE equations, and so on.
The RGE equations $know$ that $\Lambda$ is a special scale.
In order to compute whether scalar masses receive corrections of order $\Lambda$,
we must compute and solve the RGE equations for massive parameters.
The RGE for squared scalar masses have the following generic form, dictated by dimensional analysis:
\beq \frac{d}{d\ln \mu} {m}^2 = \hbox{(dimension-less couplings)} \times {m}^2.\eeq
The right-handed side in general contains scalar masses, fermion masses and cubic scalar couplings.
Without explicit computations it is clear that, 
if we set all masses to zero at any scale $\mu_0$ (for example a scale much above $\Lambda$),
all masses will remain zero at any other scale $\mu$ (for example a scale much below $\Lambda$).
No scalar mass of order $\Lambda$ is generated trough RGE evolution when the $\Lambda$ threshold is crossed.

In fact the RG-invariant scale $\Lambda$ appears when solving for the RG equations.
In models with a single scalar squared mass one has:
\begin{equation}
\label{m2run}
\frac{m^2(\mu)}{ m^2(\mu_0)}= \exp\left[ \int_{\mu_0}^{\mu} \left(\Delta_{m^2} -2 \right)\, d\ln \mu   \right]  \ , 
\end{equation}
with 
\begin{equation}
\frac{\beta_{m^2}}{m^2 (4\pi)^2}  = \frac{d \ln m^2}{ d\ln \mu} =  \Delta_{m^2} -2\ . 
\end{equation}
Here $\Delta_{m^2}$ is the quantum dimension of the mass operator.  
The non-trivial $\Lambda$ dependence is automatically encoded in the running of the various couplings entering in the above expression. 
While $\Lambda$-dependent, the renormalisation of $m^2$ is multiplicative:
the  additive renormalisation of order $\Lambda^2$ claimed in~\cite{Skiba} is absent.
This shows that these type of corrections do not introduce an explicit mass-term for the scalars despite the presence of an RG-invariant $\Lambda$. In addition,   according to our interpretation of the scale $\Lambda$ no special scheme is privileged. 

The ratio $m/\Lambda$ allows to measure deviations from IR quantum conformality when making the arbitrary choice of the bare mass of the scalar, or any other physical scale. Near IR quantum conformality can, in fact, be naturally ensured in the present framework requiring $m \ll \Lambda$ for any $\mu < \Lambda$. In other words we use $\Lambda$ as the RG-invariant meter to compare scales.

\bigskip

\label{RGEsol}

In order to make the discussion more explicit, we consider the model of section~\ref{LS+} 
and determine the RGE for the mass term operators $m_H^2 |H|^2 + m_S^2 |S|^2$
that would explicitly break scale symmetry (fermion masses still vanish because of the chiral symmetry).
Their one-loop RGE are:
\begin{align}
\label{mmH}
\beta_{m_H^2}^{(1)} &= m_H^2 \left[ g^2  \left(\frac{3}{N_c}-3 N_c\right)+4 \lambda_{H} (N_c+1)+2  N_F \tilde{y}^2+2  N_F \tilde{y}'^2\right]+2  m_S^2 \lambda_{HS} N_F^2,\\
\label{mmS}
\beta_{m_S^2}^{(1)} &= m_S^2 \left[ 4 \lambda_{S1} \left(N_F^2+1\right) + 8 \lambda_{S2} N_F+ 2  N_c y^2 + 2 {y^{\prime2}} \right] + 2 m_H^2 \lambda_{HS}  N_c.
\end{align}
The couplings evolve satisfying $g^2 \propto y^2 \propto \lambda$
to leading order in $\epsilon$ along the UV-attractive asymptotically-safe trajectory connecting the theory to the IR Gaussian fixed point. 
So the RGE for the masses reduce to independent equations for appropriate linear combinations $m^2_i$ of the squared masses.
Neglecting sub-leading terms in the limit of large $\N,N_F$ the RGE for $m_S^2$ depends only on itself:
\begin{equation}\label{eq:gammam}
\beta^{(1)}_{m^2_S} =\frac{6}{13} 
\sqrt{20 +6 \sqrt{23}} 
\,\cdot {g^2\N} m_S^2.
\end{equation}
Eq.\eq{gammam} can be integrated analytically, giving
\begin{equation}
\label{m2running}
\frac{m_S^2(\mu)}{ m_S^2(\mu_0)}=  
w^{  \frac{4 \epsilon}{19  } 
\sqrt{20 +6 \sqrt{23}} 
}
\qquad
\hbox{where}
\qquad w = \bigg[ \frac{ 1- g_*^2/g^2(\mu)}{ 1- g_*^2/g^2(\mu_0)}   \bigg]^{-\frac{171}{104\epsilon^2}}.
\end{equation}
Considering the fixed point with $\tilde{y},\lambda_{HS}\neq 0$, 
and defining the numerical constants  $\bar{\lambda}_H= \lambda_H/g^2$ and $\bar{\lambda}_{HS}=N_F \lambda_{HS}/g^2$ 
listed in table \ref{tab:FP},
eq.\,\eqref{mmH} can also be integrated and gives
\beq
\label{mH2running}
\frac{m_H^2(\mu) }{m^2_H(\mu_0) }=  w^{-\frac{\epsilon}{57}(67-104 \bar{\lambda}_H) } +\frac{286 \bar{\lambda}_{HS} \, m_S^2(\mu_0) /m^2_H(\mu_0) }{67 + 12 \sqrt{20 + 6 \sqrt{23}} - 104 \bar{\lambda}_H} \left( w^{\frac{4\epsilon}{19} \sqrt{20 + 6 \sqrt{23}}} - w^{-\frac{\epsilon}{57}(67-104 \bar{\lambda}_H) }  \right) \eeq
The factor $w$ can be explicitly written as ratios of Lambert functions using eq.\eq{gW},
and at ultra-high energies $\mu,\mu_0\gg\Lambda$  it reduces to $w\simeq\mu/\mu_0$.
The solutions to the RGE in this limit can be easily obtained substituting  constant couplings in eq.s~\eqref{mmH} and~\eqref{mmS}.
These expressions say that the various massive parameters $m_i$ acquire dimension $1 +{\cal O}(\epsilon)$ at energies above $\Lambda$.
Our results recover the fact that a massive scalar contributes to the mass of other scalars coupled to it. 
The physical ratios between different masses in general run by an infinite amount up to infinite energy: 
this can be seen as a motivation for considering theories where all masses vanish, being generated only at low energy trough
dynamical generation of vacuum expectation values or condensates.
However it does not mean that masses receive power-divergent corrections: no mass is generated trough RGE running, if masses vanish at some scale.

\subsection{Non-perturbative contributions}\label{nonpert}
Finally, we discuss now truly non-perturbative effects, which
could give corrections of order $\Lambda^2 e^{-1/C}$.
In the model of section~\ref{LS+} the couplings $C$ can be chosen to be arbitrarily small, such that non-perturbative effects, even if present, are exponentially suppressed.
In this model there are no new bound states with masses of order $\Lambda$, 
no condensates of order $\Lambda$, no new non-perturbative phenomena:
The RG invariant scale $\Lambda$ merely determines the boundary between the IR and the UV regime.

In order to make the discussion more concrete,
we discuss two special cases of non-perturbative phenomena.

First, if the fixed point is not fully IR-attractive
the quartics could run to low-energy values that violated the positivity condition of the potential,
cross the  boundary in eq.\eq{Vpos} at a scale $\mu\sim M$ exponentially smaller than $\Lambda$.
If this happens, scalars acquire vacuum expectation values and masses order $M$
trough the  Coleman-Weinberg mechanism.
Indeed various works proposed extensions of the Standard Model where the weak scale is generated in this way.

\smallskip

Next, we notice that the running of the SM Higgs quartic $\lambda_H$ in fig.~\ref{fig:SMRGE} exhibits 
a similar, but more complex pattern:
there is a $2-3\sigma$ hint that $\lambda_H$ runs negative between $E_{\rm min}\sim 10^{10}\GeV$ and  $E_{\rm max}\sim 10^{30}\GeV$, see fig.~\ref{fig:SMRGE}.
As well known, this implies a vacuum decay rate suppressed by the non-perturbative factor
$e^{-S}$, where $S=8\pi^2/3|\lambda_H(E)|$
is the action of the Fubini bounce
$h_E(r) = \sqrt{-2/\lambda}\times 2E/(1+E^2 r^2)$.
Here $r^2 = x^2+y^2+z^2 -t^2$ 
and $E$ is a free parameter with dimension of mass,
that arises because of classical scale invariance.
The non-perturbative Fubini bounce also leads to a non-perturbative correction to the Higgs mass of order
\beq \delta m_H^2 \sim \int_{E_{\rm min}}^{E_{\rm max}} E\, dE~e^{-S}.\eeq
Such non-perturbative correction to the Higgs mass is negligibly small,
given that vacuum decay is negligibly slow comparte to cosmological time-scale, namely
$e^{-S}\ll (H_0/E)^4$ where $H_0$ is the Hubble rate.

\vskip .5cm

We conclude this section by asserting that in an asymptotically safe theory featuring Higgs-like states no masses are generated along the trajectory connecting the IR Gaussian fixed point dynamics to the interacting UV safe one. The intrinsic and calculable RG invariant scale $\Lambda$ merely determines the boundary between the IR and UV conformal regimes. Furthermore at this scale no new fundamental degrees of freedom are generated. This is so since the underlying theory is described by the same fundamental degrees of freedom along the entire RG flow\footnote{Our theory respects the $a$-theorem inequality $\Delta a >0$ calculated between the UV and IR fixed point in the large $N_c$ and $N_f$ limit \cite{SanninoERG2016} and therefore  the UV and IR CFTs are distinct,
even though the underlying degrees of freedom remain the same along the RG flow. }.  When introducing explicit conformal symmetry breaking operators such as the Higgs mass term the scale $\Lambda$ allows us to ensure that deviations from the IR quantum conformal behaviour are minimal so that the physical mass  $m \ll \Lambda$ for any $\mu < \Lambda$. This can be naturally achieved in any UV and IR conformal preserving renormalization scheme.

Our results show that the claim  that asymptotically safe scalars are never naturally light~\cite{Skiba} maximally violates quantum IR (near) conformality by, de facto, elevating the RG invariant scale $\Lambda$ to the mass scale of the Higgs. 

%

\section{Conclusions}\label{concl}

In section~\ref{LS+} we extended the Litim-Sannino~\cite{SL} theory to further contain a Higgs-like scalar $H$ charged under the gauge group and with further Yukawa and quartic interactions. We showed that all couplings are governed by an IR Gaussian fixed point at low energies, and grow at short distance until a scale $\Lambda$. Above this scale the couplings enter into a rigorous asymptotically safe regime up to infinite energy, thereby avoiding Landau poles as illustrated in fig.~\ref{fig:modelRGE}.\footnote{In appendix~\ref{LSchiral}  our attempts to make the model more SM-like by adding chiral fermions have been described. 
Although we succeeded in making the gauge and the Yukawa couplings 
asymptotically safe we were unable to render the quartic couplings safe as well. Our analysis, however, does not exclude the possibility of building a SM-like chiral theory that is fully asymptotically safe. }   

\medskip


 In section~\ref{H} we investigated the perturbative and non-perturbative quantum corrections  to the scalar mass operator.
To better elucidate our main points, we determined these corrections for the calculable model of section~\ref{LS+}, serving as SM template. 
We showed that  no scalar masses of order of the transmutation scale $\Lambda$ are generated  
along the entire RG trajectory connecting the IR Gaussian fixed point to the UV interacting fixed point. 
Scalars which are massless above $\Lambda$ remain massless below $\Lambda$.
Only  exponentially small  non-perturbative corrections can contribute to scalar masses, as discussed in section~\ref{nonpert}.
This can be understood trough dimensional analysis, analogously to how we understand
that a fermion mass $M$ contributes to a scalar squared mass at order $M^2$. 
In our model interactions give quantum dimensions of order $\epsilon = N_c g^2/(4\pi)^2$,
which never gets of order unity because $g$ is perturbative. 
Thereby many powers are needed to form a squared scalar mass, which can only be  generated at non-perturbative order $\epsilon^{2/\epsilon}$.
Our couplings can be arbitrarily perturbative: for example a QED-like loop factor of order $\epsilon=1/1000$ corresponds to
$\delta m^2 \sim (1000^{-1000}\Lambda)^2$, contrary to~\cite{Skiba} who claimed that
scalar masses $m^2$ unavoidably receive quantum corrections $\delta m^2$ of the order of the transmutation scale $\Lambda^2$,
such that scalars lighter than $\Lambda$ would be unnatural.

If scalar masses are added to the theory, they receive the multiplicative 
RGE corrections of eq.~\eqref{m2run} that have been computed in section~\ref{RGEsol} in the explicit model. These yield a  non-trivial and non-quadratic $\Lambda$ dependence arising solely from the dynamics. 
The mass of Higgs-like scalars can be naturally small relative to the scale $\Lambda$ that acts, in this respect, merely as a comparison scale. Our results do not validate 
the claims made in \cite{Skiba} that, in practice by the use of ad-hoc regularization schemes, elevated the scale $\Lambda$ to the mass of the Higgs, maximally violating, at least the low energy (near) conformality of the theory. It is worth stressing that, differently from the naive estimates of\cite{Skiba},  never in our computations we had to resort to ad-hoc assumptions or expansions around one of the two fixed points since we can rigorously solve for the flow connecting the IR and UV, including the determination of the nontrivial anomalous dimensions. Our results naturally capture the correct power-law behaviour when approaching the UV interacting fixed point as already explained in sections F, G of~\cite{SL}.  
\bigskip

Our interest further resides in using these computable models to motivate new avenues for the SM near the scale $\Lambda\sim 10^{40}\GeV$, where the hypercharge coupling $g_Y$ nears its Landau pole. To this end a case-by-case investigation is needed. In particular a phenomenologically viable application to the SM case would deserve a proper dedicated study to firmly establish
what happens near the hypercharge Landau pole. 
Do couplings enter into an asymptotically safe regime, as illustrated in fig.~\ref{fig:SMRGE}? 
If yes, would the Higgs mass remain naturally small, or non-perturbative dynamics generate condensates of order $\Lambda$
that affect the Higgs mass?
Our explicit example demonstrates that such a possibility is worth exploring.

\subsection*{Acknowledgements}
This work was supported by the grant 669668 -- NEO-NAT -- ERC-AdG-2014 and the Danish National Research Foundation under the grant DNRF:90. We thank Ara Ioannisian for useful discussions. 
A.S.\ thanks Riccardo Rattazzi for discussions about dimensional analysis.
Martin Schmaltz is thanked for discussions and clarifications of his work during the 2016 CERN workshop {\it Charting the Unknown}.

\appendix

\section{Asymptotically safe models with chiral fermions?}\label{LSchiral}
{ Here we analyse basic models with chiral fermions, partially inspired by the traditional structures of grand unified theories. Our goal is not to be exhaustive but rather to guide the reader towards these extensions while exemplifying the challenges one faces in building TAS extensions of this kind.   In particular 
we  add one or more families of chiral fermions}.\footnote{ A more systematic and complementary analysis of chiral gauge theories featuring (gauged) scalars can be found in \cite{Molgaard:2016bqf}. Here both the totally asymptotically free and safe scenarios were investigated.  }
The minimal chiral anomaly-free  family is made of either
one anti-symmetric $\psi_A$
and $\N-4$ anti-fundamentals $\tilde\psi$
or of
one symmetric $\psi_S$ and $\N+4$ anti-fundamentals $\tilde\psi$ of $\SU(\N)$.
The two possibilities coincide in the limit of large $\N$, where
no other possibility exists: all higher
representations of $\SU(\N)$ with 3 or more indeces
 contribute to $\beta_g^{(1)}$ more than vectors at large $\N$, such that $\beta_g^{(1)}$ becomes large and positive. 
Spinors of $\SO(\N)$ have the same problem: $\beta_g^{(1)}$ cannot be small.

A look at the relevant $\SU(\N)$  group-theorethical factors
\beq
\begin{array}{c|ccc}
\hbox{representation}~R & \hbox{dimension }d_R  & T_R\hbox{ in $\Tr (T^a T^b) = T_R \delta^{ab}$} & C_R=d_G T_R/d_R \\ \hline
\hbox{singlet} & 1 & 0 & 0 \\
\hbox{fundamental} & \N &  1/2 & \N/2-1/2\N  \\
\hbox{adjoint} & \N^2-1 & \N & \N \\
\hbox{anti-symmetric} & \N(\N-1)/2 & (\N-2)/2&\N-1-2/\N  \\
\hbox{symmetric} & \N(\N+1)/2 & (\N+2)/2&\N+1-2/\N  \\
\end{array}
\eeq
shows that adding minimal chiral fermions cannot be
a correction which is  sub-leading in the limit of large $\N$
with respect to the  LS model.
A non-trivial feature of this model is that the Yukawa contribution makes
$\beta_g^{(2)}  = \beta_g^{(2)} |_{\rm gauge} + \beta_g^{(2)} |_{\rm Yukawa}$ negative.


In general, it is not easy to satisfy this crucial condition.
The generic expressions for the gauge beta functions
\begin{eqnsystem}{sys:betag}
\beta_g^{(1)} &=& g^3 \bigg[-\frac{11}{3} C_{G} + {\sum_F} \frac{2}{3} T_{F} + {\sum_S \kappa_S \frac{1}{3}} T_{S}\bigg],\\
\beta_g^{(2)} |_{\rm gauge} &=& g^5\bigg[ -\frac{34}{3} C_{G}^2+\sum_F \bigg(2 C_{F} +\frac{10}{3} C_{G}\bigg) T_F+
 {\sum_S \kappa_S} \bigg({4} C_S+\frac{{2}}{3} C_G \bigg)T_S \bigg]
\,,
 \end{eqnsystem}
(where $\kappa_S=1,\,1/2$ for complex or real scalars respectively)
show that, choosing a matter content such that $\beta_g^{(1)}=0$, 
$\beta_g^{(2)} |_{\rm gauge} $ is positive and minimal if the matter content consists only of fermions in the fundamental,
as in the LS model,
where the total $\beta_g^{(2)} $ is negative by a relatively small amount, see eq.\eq{betag2}.
Adding one or more chiral families and a small number ($\ll N_c \sim N_F$) of Higgs scalars in the fundamental coupled to the fermions as $H^* \tilde\psi \psi_{A,S}$, we find that the conditions $\beta_g^{(1)}=0$  and $\beta_g^{(2)}<0$ cannot be satisfied together. No perturbative UV-interacting fixed point can be found for the gauge and yukawa couplings.

\smallskip

We then need to consider more involved models.
In general, it is convenient to add as many singlet fermions and/or scalars as possible,
as they can allow for extra Yukawa couplings contributing to a negative $\beta_g^{(2)}|_{\rm Yukawa}$
without contributing to $\beta_g^{(2)} |_{\rm gauge}$. Indeed the LS model introduces many singlet scalars $S$.

Table~\ref{tab:symmetries} shows a model candidate to be Totally Asymptotically Safe, obtained adding
one chiral family to the LS model, some Higgs-like scalars $H$ and some neutral scalars $\tilde S$.
A more complex pattern of Yukawa couplings is allowed, making more complicated to compute their possible fixed points.

\begin{table}[t]\small
$$\begin{array}{|c|c|ccc|}\hline
\hbox{Fields}& \hbox{Gauge} & \multicolumn{3}{c|}{\hbox{Global symmetries}} \\
\hline  
     & \SU(\N) & \U(N_F)_L & \U(N_F)_R  & \U(\N \pm 4) \\ 
\hline  \hline
\rowcolor[cmyk]{0.1,0,0,0} 
\psi & {\tiny  \yng(1)} & \overline{\tiny  \yng(1)} & 1 & 1   \\
\rowcolor[cmyk]{0.1,0,0,0} 
\bar\psi & \overline{\tiny  \yng(1)} & 1 & {\tiny  \yng(1)} & 1   \\
\rowcolor[cmyk]{0.1,0,0,0} 
S & 1 & {\tiny  \yng(1)} & \overline{\tiny  \yng(1)} & 1 \\
\hline \hline 
\rowcolor[cmyk]{0,0.1,0,0} 
\psi_{A,S} &   {\tiny \yng(1,1)},{\tiny  \yng(2)}  & 1 &1 & 1 \\
\rowcolor[cmyk]{0,0.1,0,0} 
\tilde\psi &  \overline{\tiny  \yng(1)} & 1 &1 & {\tiny  \yng(1)} \\
\rowcolor[cmyk]{0,0.1,0,0}  
\tilde{S} & 1 & {\tiny  \yng(1)} & 1 & \overline{\tiny  \yng(1)} \\
\rowcolor[cmyk]{0,0.1,0,0}  
H & {\tiny  \yng(1)} & 1 & 1 & {\tiny  \yng(1)} \\ 
 \hline
\end{array}$$
\caption{\em\label{tab:symmetries}  A candidate model with anomaly-free chiral fermions $\tilde\psi\oplus\psi_{A,S}$ and Higgs-like scalars $H$ that might satisfy Total Asymptotic Safety.  
As in table \ref{tab:SL-chargedscalar}, the upper box represents the original LS model, while in the lower box are listed the extra fields.}
\end{table}

Fixed points correspond to specific values of the couplings such that all their beta-functions vanish.
The theories under consideration allow for $\sim \N^3$ Yukawa couplings.
In a theory with many couplings the equations $\beta=0$ can have many different solutions.
Each solution seems to correspond to a specific flavour symmetry, because
couplings with the same gauge quantum numbers have the same beta functions.
Although we don't know whether this is a generic mathematical result,
we proceed by computing the $\beta$ functions assuming the various possible flavour symmetries,
such that the number of independent Yukawa couplings is reduced to a few.
For example,  the LS model assumed the maximal flavour symmetry allowed by its
matter content, see table~\ref{tab:SL-chargedscalar}, such that there is only one independent Yukawa coupling $y$.
In table~\ref{tab:symmetries} we again assume a quasi-maximal flavour symmetry.
The most generic Yukawa interactions allowed by the gauge and global flavour symmetries then are
\beq
\label{y}
\Lag_Y= y_1 S  \psi \bar{\psi} + \tilde{y}_1 \tilde S  \psi  \tilde{\psi} + \tilde{y}_2 H^* \psi_{A,S} \tilde{\psi}+\hbox{h.c.}
\eeq
The one-loop RGE for the gauge coupling is
\beq \beta^{(1)}_g = g^3 \bigg[ - \frac{17}{6} \N + \frac{2 }{3} N_F \pm \frac83\bigg] = g^3 \frac{2 \N}{3}\epsilon ,\qquad
\epsilon = \frac{N_F \pm 4}{\N}-\frac{17}{4}. \eeq
The RGE for the Yukawa couplings in the relevant limit $\epsilon \ll 1$ and $\N \gg 1$ are
\beq  \left\{\begin{array}{l}
\beta_{y_1}^{(1)} \simeq  \N y_1 [-3 g^2 +\frac{21}{4} y_1^2+\frac12 \tilde{y}_1^2],\\
  \beta_{\tilde{y}_1}^{(1)} \simeq \frac14 \N \tilde{y}_1 [-12 g^2 +\frac{17}{2} y_1^2+\frac{29}{2} \tilde{y}_1^2+ \tilde{y}_2^2],\\
  \beta_{\tilde{y}_2}^{(1)} \simeq \frac14 \N \tilde{y}_2 [-18 g^2 +\frac{17}{2} \tilde{y}_1^2+5\tilde{y}_2^2].
\end{array}\right.\eeq
They have a few fixed-point solutions.  One solution is
\beq 
\label{FP}
y_1^2=\frac{684}{1259}g^2, \qquad \tilde{y}_1^2=\frac{372}{1259}g^2, \qquad \tilde{y}_2^2=\frac{3900}{1259}g^2.
\eeq
This fixed point leads to a negative 
\beq \beta_g^{(2)}  \stackrel{ \N\gg 1}{\simeq}\frac{53}{4}  g^5 \N^2 \bigg( 1 - \frac{67443}{66727}\bigg)\eeq
such that $g$ has a fixed point, that we compute at leading order in $\epsilon$ and $1/\N$:
\beq g_*^2 \simeq \frac{2518}{537}  \frac{(4\pi)^2 }{ \N}\epsilon.\eeq
Again, $\epsilon$ can be made arbitrarily small, such that $g_*$ can be perturbative.
The scalar potential that respects the assumed flavour symmetry
contains 11 scalar quartics:
\begin{eqnarray}
\label{lambda}
V &=& \lambda_{S} \Tr[S^\dag S]^2 + \lambda_{S}' \Tr[S^\dag S S^\dag S] + \lambda_{\tilde{S}} \Tr[\tilde{S}^\dag \tilde{S}]^2 + \lambda\_{\tilde{S}} \Tr[\tilde{S}^\dag \tilde{S} \tilde{S}^\dag \tilde{S}] +\notag \\
&&+ \lambda_{H} (H^\dag H)^2  
+\lambda'_{H} \Tr[H^\dag H H^\dag H] 
+ \lambda_{S\tilde{S}} \Tr[S^\dag S] \Tr[\tilde{S}^\dag \tilde{S}] + \lambda'_{S\tilde{S} } \Tr[S S^\dag \tilde{S} \tilde{S}^\dag] + \notag \\
&&+ \lambda_{HS} \Tr[S S^\dag] (H^\dag H)  
+ \lambda_{H \tilde{S}} \Tr[\tilde{S}\tilde{S}^\dag] (H^\dag H) + \lambda'_{H \tilde{S}} \Tr[\tilde{S} H H^\dag \tilde{S}^\dag]  \,.
\end{eqnarray}
We computed their RGE equations, but we don't find any fixed point for them.
There seems to be no particular reason: just a matter of unlucky order one factors
in a system of 11 quadratic equations that would fill a page.
Adding extra neutral fermions $N$ transforming as anti-fundamentals of $\SU(N_F)_R \otimes \SU(N_c \pm 4)$ and $N'$ transforming as fundamentals of $\SU(N_F)_L \otimes \SU(N_c \pm 4)$ allows for extra Yukawa couplings
analogous to the one in eq.\eq{LagY};
however we cannot find other fixed point solutions for the gauge and yukawa couplings apart from the solution of eq.~\eqref{FP}.

\medskip

We tried to reduce the global symmetries in order to search for more generic fixed points.
For example the symmetric and anti-symmetric components of $S$ transform
independently under the flavour symmetry $\SU(N_F)_V$,
such that $V$ now contains 14 independent quartics.
Still, they have no fixed points.
Alternatively, the global symmetry 
$\SO(N_F)_L \otimes \U(N_F)_R \otimes \SO(N_c\pm 4)$
is respected if the scalars $\tilde S$ are real, changing order one factors in the RGE.
The  allowed Yukawa couplings
\beq
\Lag_Y= y_1 S  \psi \bar{\psi} + \frac{\tilde{y}_1}{\sqrt{2}} \tilde S  \psi  \tilde{\psi} + \tilde{y}_2 H^\dagger \psi_S \tilde{\psi} \,,
\eeq
have a fixed point 
\beq y_1^2=\frac{924}{1679}g^2, \qquad \tilde{y}_1^2=\frac{744}{1679}g^2 ,\qquad \tilde{y}_2^2=\frac{5412}{1679}g^2  \,,
\eeq
corresponding to a negative $\beta_g^{(2)}$.
Again, the 13 quartics now allowed by the symmetries do not have any fixed  point.
Finally, we tried adding extra neutral fermions that allow for extra Yukawa couplings, but this does not have a beneficial effect.  We reported the negative results regarding our preliminary analysis of chiral matter to help elucidate the challenges one faces with this class of models but also to guide the reader towards more successful attempts like the ones in section V of \cite{Molgaard:2016bqf }.

\end{document}